\documentclass[traditabstract,longauth]{aa}

\usepackage{txfonts}
\usepackage{natbib}
\usepackage{graphicx}
\bibpunct{(}{)}{;}{a}{}{,}
\newcommand{\referee}[1]{\textrm{#1}}

\newcommand{\Herschel}{\textit{Herschel}}

\newcommand{\HHOp}  {\mbox{H$_2$O$^+$}}         
\newcommand{\OHp}  {\mbox{OH$^+$}}         
\newcommand{\HHO}  {\mbox{H$_2$O}}         
\newcommand{\HHHOp}  {\mbox{H$_3$O$^+$}}         
\newcommand{\emm}[1]{\ensuremath{#1}}   
\newcommand{\emr}[1]{\emm{\mathrm{#1}}} 
\newcommand{\unit}[1]{\emm{\, \emr{#1}}}

\newcommand{\pscm}{\unit{cm^{-2}}}
\newcommand{\pccm}{\unit{cm^{-3}}}

\newcommand{\kms}   {\unit{km\,s^{-1}}}

\newcommand{\eccs} {\unit{erg cm^{-3}s^{-1}}}

\begin{document}
%
%
  \title{Interstellar \OHp , \HHOp \ and \HHHOp \ along the sight-line to G10.6-0.4 
  \thanks{\Herschel \ is an ESA space observatory with science instruments provided by European-led Principal Investigator consortia and with important particiation from NASA.}
}

   \authorrunning{M. Gerin et al.}
   \titlerunning{}

\author{M.~Gerin\inst{1}, 
M.~De Luca \inst{1}, 
J.~Black\inst{2},  
J.~R.~Goicoechea \inst{3},  
E.~Herbst\inst{4}, 
D.~A.~Neufeld\inst{5}, 
E.~Falgarone \inst{1},
 B.~Godard \inst{1,8}, 
J.~C.~Pearson\inst{6},
 D.~C.~Lis \inst{7},
T.~G.~Phillips \inst{7},
T.~A.~Bell\inst{7},
 P.~Sonnentrucker\inst{5},  
 F.~Boulanger \inst{8}, 
J.~Cernicharo  \inst{3}, 
A.~Coutens \inst{12,13}
E.~Dartois \inst{8}, 
P.~Encrenaz \inst{1}, 
T.~Giesen\inst{9}, 
P.~F.~Goldsmith  \inst{6}, 
H. Gupta \inst{6}
C.~Gry \inst{10}, 
P.~Hennebelle \inst{1}, 
P.~Hily-Blant \inst{11}, 
C.~Joblin \inst{12,13} , 
M.~Kazmierczak\inst{15}
R.~Kolos  \inst{14}, 
J.~Krelowski \inst{15}, 
J.~Martin-Pintado \inst{3},  
R.~Monje \inst{7}
B.~Mookerjea \inst{15},  
M.~Perault \inst{1}, 
C.~Persson \inst{2}, 
R.~Plume \inst{16},
P.B.~~Rimmer\inst{4},
 M.~Salez \inst{1}, 
M.~Schmidt\inst{18}, 
J.~Stutzki  \inst{9}, 
D.~Teyssier\inst{19}, 
C.~Vastel \inst{12,13}, 
S.~Yu \inst{6},
A.~Contursi  \inst{20}, 
K.~Menten \inst{21},
 T.~Geballe \inst{22},
S.~Schlemmer \inst{9},
R.~Shipman \inst{23},
A.~G.~G.~M.~Tielens \inst{24},
S.~Philipp-May \inst{21,25},
A.~Cros \inst{12,13},
J.~Zmuidzinas \inst{7},
L.~A.~Samoska \inst{6},
K.~Klein \inst{25}, 
A.~Lorenzani \inst{27}
}

    \institute{
 LERMA, CNRS, Observatoire de Paris and ENS, France \\  
\email{maryvonne.gerin@ens.fr}
\and Chalmers University of Technology, G\"oteborg, Sweden.  
\and  Centro de Astrobiolog\'{\i}a, CSIC-INTA, Madrid, Spain. 
\and Depts.\ of Physics, Astronomy \& Chemistry, Ohio State Univ.,  
USA.
\and The Johns Hopkins University, Baltimore, MD 21218, USA 
\and  JPL, California Institute of Technology, Pasadena, USA.
\and California Institute of Technology, Pasadena, CA 91125, USA 
\and Institut d'Astrophysique Spatiale (IAS), Orsay, France. 
\and I. Physikalisches Institut, University of Cologne, Germany. 
\and Laboratoire d'Astrophysique de Marseille (LAM), France. 
\and Laboratoire d'Astrophysique de Grenoble, France. 
\and  Universit\'e de Toulouse ; UPS ; CESR ; 9 avenue du colonel Roche,
F-31028 Toulouse cedex 4, France
\and CNRS ; UMR5187 ; F-31028 Toulouse, France 
\and Inst.\ of Physical Chemistry, Polish Academy of Sciences, Poland 
\and Nicolaus Copernicus University, Toru´n, Poland. 
\and Tata Institute of Fundamental Research, Mumbai, India.
\and Dept. of Physics \& Astronomy, University of Calgary, Canada.
\and  Nicolaus Copernicus Astronomical Center, Poland 
\and European Space Astronomy Centre, ESA, Madrid, Spain. 
\and MPI f\"ur Extraterrestrische Physik, Garching, Germany. 
\and MPI f\"ur Radioastronomie, Bonn, Germany. 
\and Gemini telescope, Hilo, Hawaii, USA 
\%and Nicolaus Copernicus Astronomical Center, Poland
\and SRON Netherlands Institute for Space Research, Netherlands 
\and Sterrewacht Leiden, Netherlands 
\and Deutsches Zentrum f\"ur Luft- und Raumfahrt e. V., Raumfahrt-Agentur, 
Bonn, Germany. 
\and Department of Physics and Astronomy, University of Waterloo, Canada
\and Osservatorio Astrofisico di Arcetri-INAF- Florence - Italy 
}

\date{Received / Accepted }

%
%
\abstract
{We report the detection of absorption lines by the reactive ions
\OHp, \HHOp and \HHHOp\ along the line of sight to the submillimeter continuum 
source G10.6$-$0.4 (W31C). We  used the \Herschel\ HIFI instrument in 
dual beam switch mode to observe the ground state rotational transitions of 
\OHp\ at 971~GHz,  \HHOp\ at 1115 and 607~GHz, and \HHHOp\ at 984 GHz. 
The resultant spectra show deep absorption over a broad velocity range that
originates in the interstellar matter along the line of sight 
to  G10.6$-$0.4 as well as in the molecular gas directly 
associated with that source.
The \OHp\ spectrum reaches saturation over most velocities 
corresponding to the foreground gas, while the opacity of the
\HHOp\ lines remains lower than 1 in the same velocity range, and
the \HHHOp line shows only weak absorption.
For Local Standard of Rest (LSR) velocities between 7 and 50 kms$^{-1}$ we estimate 
total column densities  of $N$(\OHp) $\geq 2.5 \times 10^{14}$ cm$^{-2}$, 
$N$(\HHOp) $\sim 6 \times 10^{13}$ cm$^{-2}$ and 
$N$(\HHHOp) $\sim 4.0 \times 10^{13}$  cm$^{-2}$.  
These detections confirm the role of O$^+$ and OH$^+$ in initiating the oxygen
chemistry in diffuse molecular gas \referee{and strengthen our understanding of the gas phase production of water}. The high ratio of the \OHp\ by the \HHOp\ 
column density implies that these species 
predominantly  trace low-density gas with a small fraction of 
hydrogen in molecular form.  

}

%
%

\keywords{{ISM: abundances} - {ISM: molecules}}

\maketitle

\label{firstpage}

%
%
\section{Introduction}
Oxygen chemistry in the interstellar medium is initiated by
cosmic-ray or X-ray ionization of hydrogen. 
The resulting H$^+$ and H$_3^+$
transfer their charge to form the O$^+$ and OH$^+$ ions, both of which react
rapidly with H$_2$ molecules. Subsequent H-atom abstraction reactions
proceed to the terminal ion H$_3$O$^+$ with high efficiency, unless the
electron fraction is unusually high. H$_3$O$^+$ does not react with H$_2$
or any other abundant neutral species, but is destroyed by dissociative
recombination with electrons. Recombination produces O, OH, and H$_2$O.
\referee{This sequence is the main gas-phase chemical route to water in the
interstellar medium.} 
Until now, this classical scheme of the oxygen chemistry (Herbst \&
Klemperer 1973) has been tested mainly through observations of OH and 
H$_2$O, whose abundances are affected by many other processes as well, 
and by a few observations of H$_3$O$^+$. With the HIFI instrument on the
\Herschel \ satellite it is now possible to observe the strongest 
transitions of light hydrides, including the reactive molecular ions
OH$^+$ and H$_2$O$^+$ that are thought to play crucial intermediate 
roles in the formation of OH and H$_2$O.

\referee{Previous attempts to observe interstellar  H$_2$O$^+$ and OH$^+$ 
from optical spectroscopy are
very limited. Smith et al. (1984) report an 
upper limit of $10^{12}$ cm$^{-2}$ for  H$_2$O$^+$ based on optical absorption spectroscopy. No observation of the OH$^+$ transition at  358.4 nm has
been reported until very recently (Krelowski et al., private communication). 
Submillimeter telescopes have been more successful in detecting these ions.} 
With the Atacama Pathfinder EXperiment (APEX) telescope 
Wyrowski et al. (2010)  
 detected two hyperfine-structure components of the OH$^+$ 
$N=1-0$ $J=0-1$ transition near 909.2 GHz towards the strong continuum
source Sgr B2(M) and concluded that \OHp \ plays a major role in the chemistry of the diffuse molecular gas.
Using HIFI on \Herschel,  Ossenkopf et al. (2010) report
 the detection of o-\HHOp \
absorption in the three massive star-forming regions NGC~6334, 
Sgr~B2, and DR~21.

We present observations toward the massive star-forming
region G10.6$-$0.4 (W~31C) located at $RA=18h10m28.70s$, $Dec =-19^\circ55'50 s$ (J2000). The submillimeter-wave 
continuum emission is associated with a cluster of ultra-compact
H\,{\sc ii}
regions located in the so-called ``30 km s$^{-1}$ arm'' at a kinematic
distance of 4.8 kpc (Fish et al. 2003, Corbel \&\ Eikenberry 2004). 
The molecular cloud surrounding the star-forming complex shows evidence
of continuing accretion onto the embedded stars (Keto \&\ Wood 2006; 
Sollins \&\ Ho 2005). As discussed by Neufeld et al. (2010),
line absorption by foreground gas has been detected from [OI]
(Keene et al. 1999), $^{13}$CH$^+$ (Falgarone et al. 2005), HCO$^+$, 
HCN, HNC, CN (Godard et al. 2010), and atomic hydrogen (Fish et al. 2003).
While the gas associated directly with G10.6$-$0.4 is detected in
the velocity range $-5$ to $+5$ km s$^{-1}$, the foreground gas 
along the line of sight is confined to the velocity interval from 
$\sim 5$ to 45 km s$^{-1}$.

\section{Spectroscopy}

\begin{table}
\caption{Transition spectroscopic parameters}
\label{tab:spec}
\begin{tabular}{lccccl}
\hline \hline
Transition & Frequency & Error & E$_l$  & A & Ref \\
&   MHz      & MHz    & cm$^{-1}$  &  10$^{-2}$s$^{-1}$ & \\
\hline
\OHp $N = 1- 0$ \\
 $2,5/2 - 1,3/2$ & 971803.8  & 1.5 & 0.0  & 1.82& 1 \\
 $2,3/2 - 1,1/2$ & 971805.3  & 1.5 & 0.0  & 1.52& 1 \\
 $2,3/2 - 1,3/2$ & 971919.2  & 1.0 & 0.0  & 0.30& 1 \\
\hline
o-\HHOp $1_{1,1}-0_{0,0}$\\
 $3/2,3/2 - 1/2,1/2$ &  1115122.0& 10&0.0  &1.71   & 2 \\
 $3/2,1/2 - 1/2,1/2$ &  1115158.0 &10&0.0  & 2.75 & 2\\
 $3/2,5/2- 1/2,3/2$ & 1115175.8 & 10&0.0  & 3.10 & 2\\
 $3/2,3/2 - 1/2,3/2$ &  1115235.6& 10&0.0 & 1.39& 2\\
 $3/2,1/2 - 1/2,3/2$ &  1115271.6& 10&0.0  & 0.35& 2\\
\hline
p-\HHOp  $1_{1,0}-1_{0,1}$\\
 $3/2,3/2 - 3/2,3/2$ & 607207.0 & 20 & 20.9  & 
0.60 &2 \\
\hline
\HHHOp\\
$0_{0}^{-} - 1_{0}^{+}$ & 984711.9 & 0.3 &  5.1 & 2.3 & 3\\
\hline
\end{tabular}

1, M\"uller et al. 2005 \&  CDMS; 2, Strahan et al. 1986, Ossenkopf et al. 2010; 3 Yu et al. 2009 \& JPL catalog
\end{table}

%
%
%
%
Table \ref{tab:spec} summarizes the spectroscopic parameters used
for the searched transitions. 
The \OHp\ ion has a $^3$$\Sigma^-$ electronic ground state. We used
the data from  Beekoy et al. (1985) 
with reported uncertainties for the 
line frequencies better than about 1.5~MHz. M\"uller et al (2005)
have published a global fit using these data 
which is available in the Cologne Data base
for Molecular Spectroscopy (CDMS). The ground state dipole moment
obtained from theoretical calculations is 2.3 D (Werner et al. 1983, Cheng et
al. 2007).

The spectroscopic properties of \HHOp \ are extensively
discussed in the accompanying letter (Ossenkopf et al. 2010).
Its dipole moment is very similar to that of \OHp, as recently
calculated by Wu et al. (2004), who found the ground state 
dipole moment to be 0.933, mistakenly printed in atomic units 
for a value of 2.370 Debye.  
This molecular ion has ortho and para symmetry states, which have the
opposite parity to the water vapor molecule, i.e. the lowest state is the
ortho state. Therefore the ortho/para ratio should be at least 3:1.

%
%
%
%
\HHHOp \ has been the subject of many spectroscopic investigations, because for  the frequency of the inversion transition of this molecular ion
is comparable to 
the rotational constant. The most recent analysis has been presented by
Yu et al. (2009) and is available in the JPL spectral line catalog.  
We observed the lowest transition from the E symmetry state, which is $\sim
5$~cm$^{-1}$ above the ground state.

%
%
\section{Observations}
%
%

The observations were carried out by the HIFI instrument on board 
\Herschel \ (Pilbratt et al. 2010; de Graauw et al. 2010), in
the double beam switching (DBS) mode. We used 
the Wide Band Spectrometer (WBS) with 4GHz bandwidth and 
1.1 MHz effective spectral resolution. 
To help separate the signals from the
lower and upper sidebands, three integrations were carried out at
each frequency, with the local oscillator (LO) shifted by 15 kms$^{-1}$.

The data were first processed with HIPE \citep{ott} and subsequently 
exported to
CLASS. We checked that the target lines did not suffer from any
contamination by lines from the other sideband \referee{and that the absorption features were not contaminated by emission in the reference beam}. 
The two polarizations excellently agree
---the continuum levels agree to better than
10\%---except for the OH$^+$ line, where the V polarization shows
ripples and has therefore not been used in the present analysis. We
combined the data from the three LO tunings and the two polarizations to
obtain the average spectra shown in Fig. \ref{fig:spec}. 
For double-side band mixers with equal sideband gains, fully saturated 
lines are expected to absorb only half of
the continuum signal. This expectation is borne out in the present data. 
While the \OHp \ line is clearly saturated, both \HHOp lines show moderate
absorption, while the \HHHOp \ line only shows weak absorption. 

%
%
\section{Results}
%
%

\begin{figure}
\includegraphics[width=0.35\textwidth]{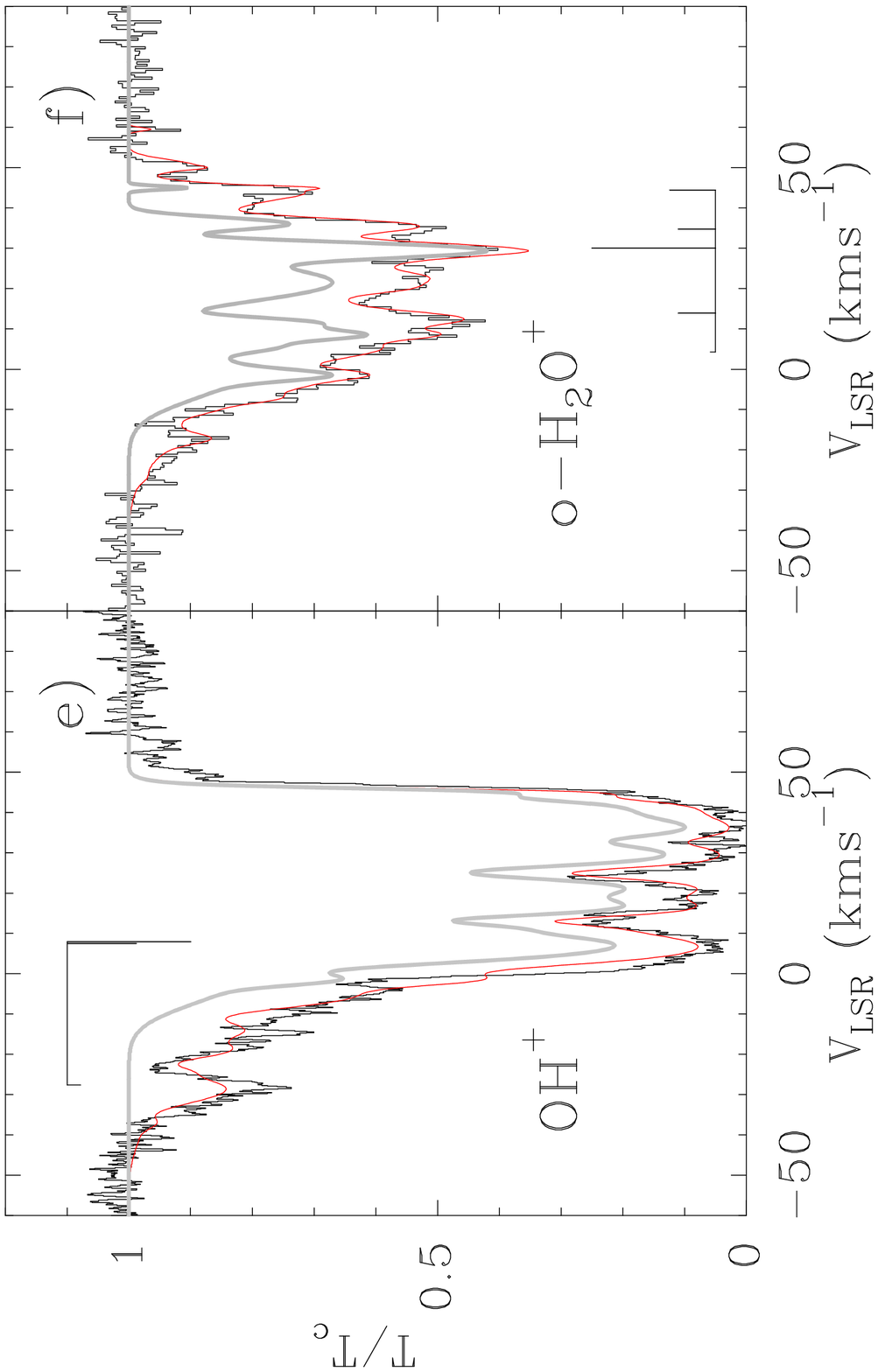}
\rotatebox{270}{
\includegraphics[width=0.22\textwidth]{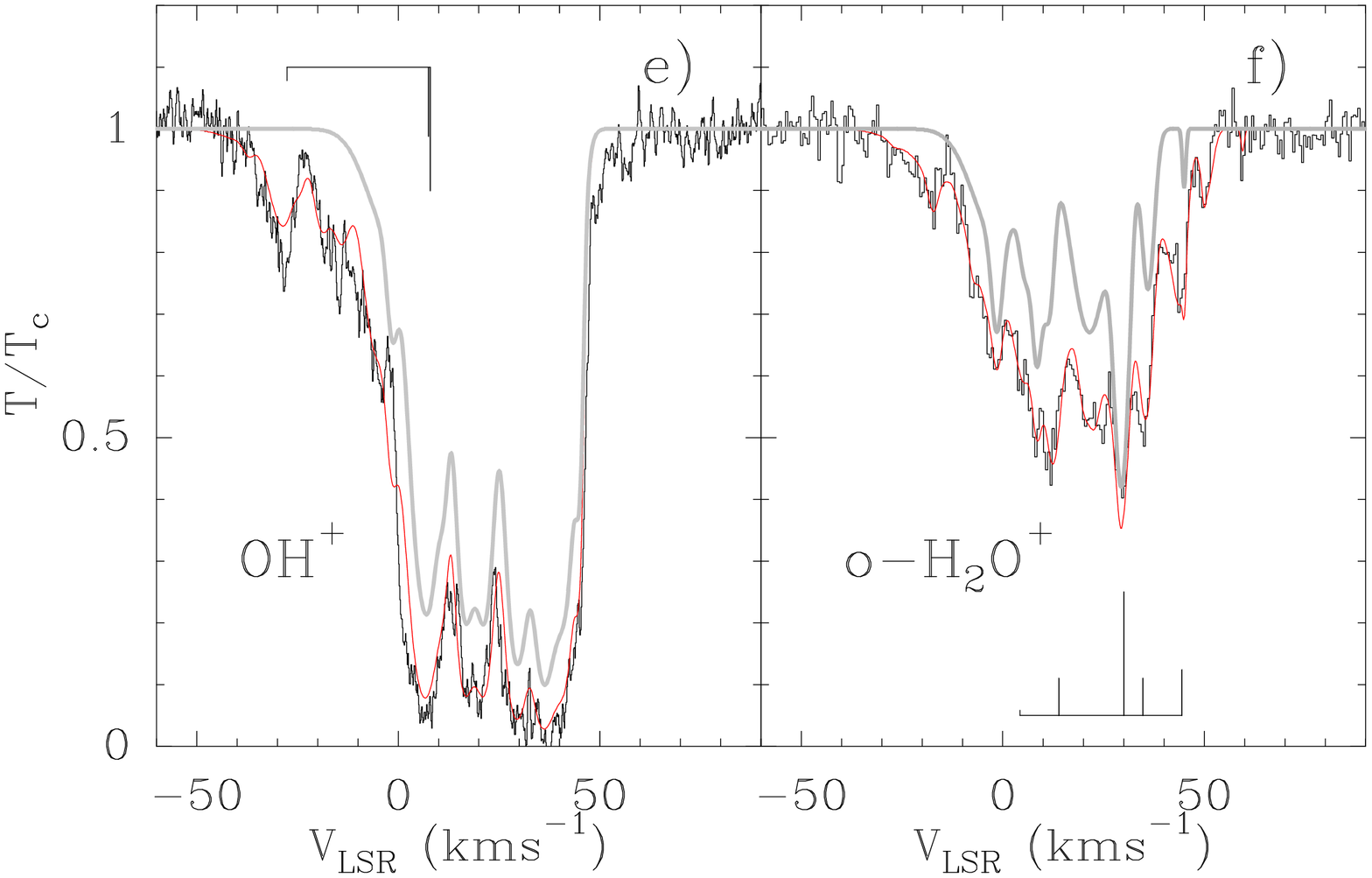}
}
\caption{HIFI spectra of (a) OH$^+$, (b) o-H$_2$O$^+$,
(c) p-H$_2$O$^+$  and (d) \HHHOp. The hyperfine structure of
the \OHp\ and o-\HHOp\ lines (see Table \ref{tab:spec}) is indicated. 
In both cases, the velocity scale corresponds to 
 the rest frequency of the strongest hyperfine component.
In panel c), the velocity scale refers to the frequency of 
the H$^{13}$CO$^+$ (6-5) line at
607174.65~MHz. Panel d) uses the frequency of 984711.907~MHz. 
 Panels e) and f)  show overlays with
the synthetic spectra   constructed with the empirical models detailed
in Table \ref{tab:ohp}. Grey curves show the profile of the strongest
hyperfine component. }
\label{fig:spec}
\end{figure}

To understand the structure of the absorption profiles and
derive  column densities of the reactive ions, we normalized
the spectra by the SSB continuum intensities. 
We  constructed empirical models of the line profiles by
combining multiple velocity components, defined by their
central velocity, line width, and opacity. The predicted 
spectrum was then computed by
combining the contributions of the hyperfine components (3 for \OHp \ and
5 for \HHOp \ as listed in Table \ref{tab:spec}) to the overall line profile. 
We show in Fig. \ref{fig:spec}  spectra
normalized by the continuum intensities, with the synthetic spectra
 superimposed. Table \ref{tab:ohp}  lists the model parameters. 

Although we obtained good fits of the \OHp \ and \HHOp\ profiles,  
it is likely that the solution is not unique because the number of velocity
components is not well constrained. 
The \OHp \ ground state lines are known with a good accuracy of $\sim 1.5$ MHz.
Velocities obtained by the empirical model agree well  with the
velocities where deep HCO$^+$(1-0) (Keene et al. 1999, Godard et
al. 2010), p-H$_2$O and HF (Neufeld et al. 2010) and HI \citep{Fish03}
absorptions are detected. It is remarkable that one of the strongest \OHp \
absorption features at V$_{LSR} \sim 7$ \kms coincides with a strong dip in the
HI profile that has no counterpart in CO emission \cite{Falgarone05}. 
Overall, the velocities of the \OHp\ and \HHOp\ velocity components agree
well. 
Although the lines are detected in absorption,  given the similarity
of their  Local Standard of Rest (LSR) velocities with the rest velocity 
of G10.6$-$0.4, it is
likely that the most blueshifted velocity components 
 are associated with the G10.6$-$0.4 molecular
complex.

Below we discuss the column densities and abundances of
\OHp, \HHOp, and \HHHOp\ in the foreground interstellar gas, with
LSR velocities higher than 5 \kms. Neufeld et al. (2010) discuss the total column
density of this  material and obtain a column density of hydrogen 
in all forms of $N_{\rm H} \sim 2.7 \times 10^{22}$~cm$^{-2}$. This figure is 
consistent with the lower limit derived from the oxygen column density measured
with ISO, $5 \times 10^{18}$~cm$^{-2}$ (Keene et al. 1999), which gives
$N_{\rm H} > 1.7 \times 10^{22}$ cm$^{-2}$ using ${\rm O/H} = 3 \times 10^{-4}$.
For this line of sight the estimated atomic hydrogen column density is
at most $N({\rm HI}) =  1.2 \times 10^{22}$~cm$^{-2}$ (Godard et al. 2010),
 hence the average
 fraction of hydrogen in molecular form is $f(H_2) \sim 0.6$, with significant
variations among the velocity components.

The total \OHp \ 
and \HHOp \ column densities were obtained by summing the contributions
of the velocity components with centroids larger than 7~kms$^{-1}$. 
We derive $N$(\OHp) $\geq 2.5 \times 10^{14}$~cm$^{-2}$ and $N$(o-\HHOp) = $4.4
\times10^{13} $~cm$^{-2}$. Although the absorption from the
p-\HHOp \ line at 607 GHz is blended with emission from H$^{13}$CO$^+$(6-5) at 607175 MHz, weak absorption is clearly detected in the velocity range of the
foreground gas along the line of sight to G10.6$-$0.4. The integrated opacity
for the LSR velocity range 20  to 45~km\,s$^{-1}$ is $\int \tau d\rm{v} = 2  kms^{-1}$, 
corresponding to a total column density of $\sim 1 \times 10^{13}$ cm$^{-2}$.
This figure is obtained for a limited velocity range, and should therefore
be considered a lower limit. Given the uncertainties, the data are 
consistent with an o/p ratio of $3$ for \HHOp, as
expected from the spin statistics. The total column density of \HHOp \
is therefore $N$(\HHOp) $\sim 6 \times 10^{13}$ cm$^{-2}$ and the 
abundance ratio $N$(\OHp)/$N$(\HHOp) $\geq 4$.
 Using the total gas column density , we conclude that \OHp \ and \HHOp \ reach fractional abundances, of
$N(\OHp)/N_{\rm H}  \geq  \times 10^{-8}$ and 
$N(\HHOp)/N_{\rm H}  \sim 2 \times 10^{-9}$.

\begin{table}
\caption{Empirical models used to analyze the \OHp \ and \HHOp  spectra}
\label{tab:ohp}
\begin{tabular}{l|rcl|rcl}
\hline \hline
         &        &    \OHp  &               &  &       \HHOp  & \\
V$_{LSR}$ & $\Delta_V$ & $\tau$ & N  & $\Delta_V$ & $\tau$ & N\\
  kms$^{-1}$ & kms$^{-1}$  & & 10$^{13}$ cm$^{-2}$ & kms$^{-1}$  & & 10$^{12}$ cm$^{-2}$  \\ 
\hline
-1.0 & 14.0 & 0.2 & 1.3  & 14.0 & 0.2 & 12\\
-1.5 & 3.0 & 0.2 & 0.3  &   3.0    & 0.2 & 2.5\\
5.5  & 6.0 &   1.0 &2.8   & 3.0  &    0.15 & 1.9 \\
8.5 & 5.0 & 0.8 & 1.9  & 3 & 0.4 & 5.0\\
11.5 & 3.0  &  0.5 & 0.7   &  3.0 &    0.3 &3.8  \\
16.5 & 5.0 &   1.5 & 3.5     &     &  &  \\
21.5 & 5.0 &   1.5 & 3.5    &  10.0   & 0.4 & 1.7 \\
29.5 & 6.0 &   2.0 & 5.6   & 4.0  &  0.8 & 1.3\\
36.0 & 5.0 &   2.0 & 4.7     &3.5  &  0.3 & 4.4\\
41.0 & 6.0 &   1.5 & 4.2      & &    & \\
45.0 & 2.0 & 0.5 & 0.5  & 1.0 & 0.1 & 0.4\\
\hline
\end{tabular}

The \OHp \ and \HHOp \ parameters refer to the frequency of the strongest 
hyperfine component \referee{(cf Table \ref{tab:spec})}.
\end{table}

\begin{figure}
\centering
\rotatebox{270}{
\includegraphics[width=0.25\textwidth]{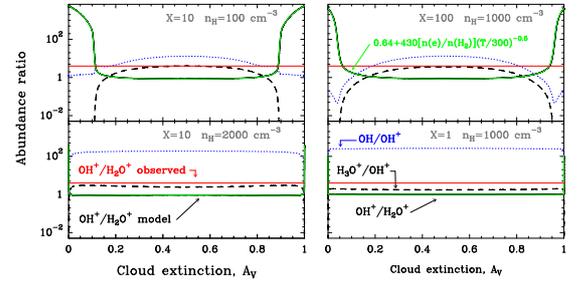}}
\caption{Predictions of the Meudon-PDR model for a slab of 
$A_v$ = 1 mag illuminated from both sides. The radiation field and 
gas density are
given in each panel. Full lines: n(\OHp)/n(\HHOp), 
dotted lines: n(OH)/n(\OHp), dashed lines : n(\HHHOp)/n(\OHp).
Eq. (2) determined for each Av in the model is shown \referee{in} green.
 } 
\label{fig:pdr}
\end{figure}

The \HHHOp spectrum shows a sharp feature with strongest absorption
at V$_{LSR}$ $\sim 40$~\kms and a broad component centered at V$_{LSR}$ $\sim
20$~\kms. This velocity structure is more similar to the  HF or 
p-\HHO \ spectra
than the \OHp \ and \HHOp \ velocity structure, confirming that these 
molecular ions trace different physical conditions.
We derived the \HHHOp \
column density of the foreground gas  for all 
velocity components.  Assuming an excitation temperature of 5~K, appropriate
for the  conditions of the gas in the inner galaxy, 
we derive  $N$(\HHHOp) $ \sim 4 \times 10^{13}$ cm$^{-2}$ for the LSR 
velocity range 7 -- 45 kms$^{-1}$. Therefore N(\OHp)/N(\HHHOp) $> 6$ and 
N(\HHOp)/N(\HHHOp) $\sim 1.5 $. These numbers represent the average 
for all velocity components but, given the
different velocity structure of \HHHOp\ and \OHp, the column density ratio
for individual velocity components can significantly  differ from the average
value.

%
%
\section{Chemistry}
%
%

%
%
\subsection{PDR models}
%
%

In order to  model the OH$^+$ and H$_2$O$^+$ chemistry
in diffuse clouds, we  used the Meudon PDR
code\footnote{See \texttt{http://pdr.obspm.fr/PDRcode.html}},
a photochemical model of a plane parallel and stationary PDR
(Le Petit et al. 2006, Goicoechea \& Le Bourlot 2007). 
\referee{We present models for a representative cloud of A$_V$=1 mag illuminated 
from both sides. Although the total extinction along this line of sight
is $\sim 14 mag$ (Corbel \& Eikenberry 2004, Neufeld et al. 2010), 
the line profile clearly shows  multiple 
 velocity components, revealing  interstellar clouds 
along the line of sight. As discussed by Godard et al (2010), 
each of these clouds
has low to moderate extinctions, corresponding to  diffuse or 
transluscent matter.   }

\referee{The Meudon PDR code computes the gas temperature gradient } 
by solving the
thermal balance (see Le Petit et al , 2006).
Our pure gas-phase chemical network is based on a modified version of the
Ohio State University (OSU) gas-phase network updated for photochemical
studies. The extinction of UV and visible radiation 
is computed using a dust grain size distribution following a power-law
(Mathis et al. 1977). The adopted oxygen and carbon elemental
abundances are 3.02$\times$10$^{-4}$ and 1.38$\times$10$^{-4}$, respectively.
As a starting point we used an ionization rate (of H) 
due to cosmic rays of 
$\zeta_H =  2.3 \times$10$^{-17}$ s$^{-1}$.

The oxygen chemistry starts with the
O($^3$P$_J$) + H$^+$ $\rightarrow$ O$^+$($^4$S) + H  charge transfer process,
where reaction rates are sensitive to the oxygen fine-structure level
populations
(Stancil et al. 1999, and references therein). Indeed the charge transfer
proceeds quickly when the  O($^3$P$_1$) level ($\Delta$$E$$_{12}$/$k$=228\,K)
is significantly populated, but it can be a slow proceess in cold
gas where only the  O($^3$P$_2$) ground-state level is populated
(Spirko et al. 2003). As listed in Table \ref{tab:chim},
OH$^+$ is formed by O$^+$ + H$_2$ with additional contribution from 
OH photoionization at the cloud edges, and by
O + H$_{3}^{+}$.
Therefore, the OH$^+$ abundance may be sensitive to gas temperature,
far ultraviolet (FUV) radiation field and \referee{$\zeta_H$}.

In order to investigate the possible sources of the high OH$^+$/H$_2$O$^+$
abundance ratio, Fig. \ref{fig:pdr} shows photochemical 
models for different densities, and radiation fields 
(thus different gas temperatures). High \OHp/\HHOp \ abundance ratios 
are found in the warm gas at low
extinction, where the fraction of hydrogen in  molecular form is significantly
smaller than 1.  As a limited set reactions need to be considered 
 we built a simplified analytical model \referee{based on the network 
listed in Table \ref{tab:chim}}, and find

\begin{table}

\caption{\label{tab:chim}Main chemical reactions }
\begin{tabular}{llc}
Number & Reaction & Rate (cm$^{-3}$s$^{-1}$)\\
\hline
\hline

1 & $ \rm{OH}^+ + \rm{H}_2 \rightarrow \rm{H}_2\rm{O}^+ + H $ & $k_1$ =  $1.01 \times
       10^{-9}$ \\
2 & $ \rm{H}_2\rm{O}^+ + \rm{H}_2 \rightarrow \rm{H}_3\rm{O}^+ + H$ & $k_2$ = $6.4
       \times 10^{-10}$  \\
3 & $ \rm{H}_2\rm{O}^+ + e^- \rightarrow products$ & $k_3$ = $4.3 \times 10^{-7}
       (T/300K)^{-0.5}$ \\
4 & $ \rm{H}_3\rm{O}^+ + e^- \rightarrow products$ &  $k_4$ = $4.3 \times 10^{-7}
       (T/300K)^{-0.5}$ \\
5 & $\rm{OH}^+ + e^- \rightarrow products$ & $k_5$ = $3.75 \times 10^{-8} (T/300)^{-0.5}$ \\
\hline
\end{tabular}
\end{table}

\begin{equation}
n(\OHp)/n(\HHOp) =  (k_2/k_1) + (k_3/k_1)[n(e^-)/n(H_2)]\end{equation}
\begin{equation}
= 0.64 + 430 \times(T/300)^{-0.5} \times [n(e^-)/n(H_2)] \  .
\end{equation}

\noindent 
Using $f(H_2) = 2n(H_2)/(n(H)+2n(H_2)) = 2n(H_2)/n_H)$  and assuming 
that all carbon is  ionised ($n(e^-) = n_C = 1.4 \times 10^{-4} n_H$)
Eq. (2) leads to $ n(\OHp)/n(\HHOp) = 0.64 + 0.12 \times(T/300)^{-0.5} / f(H_2)   $.
The data suggest that $N(\OHp)/N(\HHOp) > 4$, implying that $f(H_2) < 0.06$ 
at T = 100~K, an appropriate temperature for  diffuse gas in the molecular ring.
Equation (2) determined for each Av in the model is shown as green 
lines in Fig. \ref{fig:pdr}. It fits the modeled
abundance perfectly, demonstrating that the main ingredients
have been captured by this simple model. \referee{Note that the reactions 
were studied in the laboratory and their rates are known with a good accuracy.}


\subsection{The role of cosmic rays}

Like H$_3^+$ (Indriolo et al. 2007), \OHp \ and \HHOp \ can be used to
trace the cosmic ray ionization rate for H atoms $\zeta_H$. 
In the moderately warm gas 
where they  are produced, the charge exchange reaction is rapid. Then, every cosmic ray ionization of H can lead to O$^+$ and we can use the measured \OHp\ abundance
to set limits on $\zeta_H$. In addition to the reaction with H$_2$, \OHp \ is destroyed in dissociative recombination reactions with
electrons.  This implies $\zeta_H > X(\OHp)x(e^-)n({\rm H})k_5$, 
$\zeta_H > 9.2 \times 10^{-12} \times n(\OHp)$ at T=100~K.

The lower limit on the \OHp\ abundance in the atomic gas is $n(\OHp)/n(H) > 
2 \times 10^{-8}$, which places a robust lower limit of $\zeta_H >  
1.8 \times 10^{-19} n(H)$.
   This lower limit is  very conservative: depending on the 
conditions in the absorbing material, higher values of $\zeta_H$ 
 may prove necessary to account for the observed column densities 
of \OHp\ and \HHOp.  \referee{The H$_3^+$ observations of diffuse 
interstellar clouds point toward an
enhanced ionization rate at low extinction. In a future analysis 
we will consider how \OHp\ and \HHOp\ can provide additional constraints 
upon $\zeta_H$.}

\subsection{The role of turbulence}

In the warm chemistry driven by intermittent turbulent dissipation
(Godard et al.  2009), the production of OH$^+$ and
H$_2$O$^+$ is enhanced compared to UV-driven chemistry, with an efficiency 
depending only on the turbulent dissipation
rate, $\epsilon$. For  $\epsilon=2\times 10^{-24}$ \eccs, 
ten times higher than the average Galactic value  appropriate for the
inner Galaxy, the TDR model predicts $N$(CH$^+$)$= 4.3\times 10^{13}$ \pccm, 
$N$(OH$^+$)=$2 \times 10^{12}$\pscm\  and
$N$(H$_2$O$^+$)=$2.2 \times 10^{12}$\pscm\ per magnitude of gas of density 
$n_H=50$\pccm. The ratios between these three species weakly
depend on $\epsilon$. With the observed CH$^+$ column density, $6.7\times
10^{14}$ \pscm\ , derived from $^{13}$CH$^+$ (Falgarone et al. 2005)\footnote{In that
paper it was underestimated by a factor 8 because of an unjustified
approximation made in the derivation of the opacity.}), 
we therefore expect the following contributions of the
TDR chemistry in that line of sight: $N$(OH$^+$)=$3.1\times 10^{13}$\pscm, and
$N$(H$_2$O$^+$)$=3.4\times 10^{13}$\pscm.  The OH$^+$ column is well below the
observed value, while that of H$_2$O$^+$ is interestingly close.

%
%
\section{Conclusion}
%
%
\referee{The detection of interstellar \OHp, \HHOp, and \HHHOp provides
a strong support to our understanding of the gas phase oxygen chemistry,
in particular the gas phase route to water vapor.  
The combined analysis of the \OHp\ and \HHOp\  ground state}
spectra has shown that they mostly reside 
in warmish atomic gas where they are  produced in a reaction sequence initiated
by atomic H ionization 
followed by rapid reaction with traces of H$_2$.
 Their abundances and abundance ratio can be
used to constrain the fraction of hydrogen in molecular form, and to set 
a lower limit for the cosmic ray ionization in the diffuse gas, where
few other diagnostics are applicable. As these reactive molecular ions 
 can be observed from space with the \Herschel \ Space Observatory,
and in exceptional conditions from the ground, these preliminary 
conclusions should be verified by more extensive observations.
It is also expected that strong features from both \OHp \ and \HHOp\ 
will be detected in the spectra of distant galaxies, where elevated 
fluxes of ionizing radiation are produced by active nuclei and 
intense star-formation activity.

%
%
\begin{acknowledgements}
%
%
HIFI has been designed and built by a consortium of institutes and university
departments from across Europe, Canada and the United States under the
leadership of SRON, Netherlands Institute for Space Research, Groningen, The
Netherlands, and with major contributions from Germany, France and the
US. Consortium members are : Canada: CSA, U. Waterloo; France : CESR, LAB,
LERMA, IRAM; Germany : KOSMA, MPIfR, MPS; Ireland : NUI Maynooth; Italy : ASI,
IFSI-INAF, Osservatorio Astrofisico di Arcetri-INAF; Netherlands : SRON, TUD;
Poland : CAMK, CBK; Spain : Observatorio Astron\`omico Nacional (IGN), Centro
de Astrobiologia; Sweden : Chalmers University of Technology - MC2, RSS \&
GARD, Onsala Space Observatory, Swedish National Space Board, Stockholm
University - Stockholm Observatory; Switzerland : ETH Zurich, FHNW; USA :
CalTech, JPL, NHSC.  
MG, EF, MDL acknowledge the support from CNES, and ANR 
through the SCHISM project (ANR-09-BLAN-231). M.S. acknowledges support from grant N203393334 (MNiSW). 

\end{acknowledgements}

%
%

\end{document}